\documentclass[preprint2]{aastex}

\begin{document}

\title{The Luminosity Profiles of Brightest Cluster Galaxies}

\author{C.J. Donzelli, H. Muriel}

\affil{IATE, Observatorio Astron\'omico OAC, Laprida 854, X5000BGR, C\'ordoba, Argentina\\
Consejo Nacional de Investigaciones Cient\'ificas y T\'ecnicas (CONICET),
Avenida Rivadavia 1917, C1033AAJ, Buenos Aires, Argentina}

\author{J. P. Madrid}
\affil{Centre for Astrophysics and Supercomputing, Swinburne University of
Technology, Hawthorn, VIC 3122, Australia}

\begin{abstract}

We have derived detailed R band luminosity profiles and structural parameters for a total of 430 
brightest cluster galaxies (BCGs), down to a limiting surface brightness of 24.5 $mag arcsec^{-2}$. 
Light profiles were initially fitted with a S\'ersic's $R^{1/n}$ model, but  we found 
that 205 ($\sim48\%$) BCGs require a double component model to accurately match 
their light profiles. The best fit for these 205 galaxies is an inner S\'ersic model, with indices 
$n\sim1-7$, plus an outer exponential component. 

Thus, we establish the existence of two categories of the BCGs luminosity profiles: {\it single} 
and {\it double} component profiles. We found that double profile BCGs are brighter ($\sim0.2$ mag) 
than single profile BCG. In fact, the Kolmogorov-Smirnov test applied to these subsamples indicates 
that they have different total 
magnitude  distributions, with mean values $M_R=-23.8\pm0.6$ mag for single profile BCGs and 
$M_R=-24.0\pm0.5$ mag for double profile BCGs. We find that partial luminosities for both subsamples 
are indistinguishable up to $r = 15$ kpc, while for $r > 20$ kpc the luminosities we obtain are on 
average 0.2 mag brighter for double profile BCGs. This result indicates that extra-light for double 
profile BCGs does not come from the inner region but from the outer regions of these galaxies.
  
The best fit slope of the Kormendy relation for the whole sample is $a = 3.13\pm0.04$. However,
when fitted separately, single and double profile BCGs show different slopes: $a_{single} = 3.29\pm0.06$
and $a_{double}= 2.79\pm0.08$. Also,  the logarithmic slope of the metric luminosity $\alpha$ 
is higher in double profile BCGs ($\alpha_{double} = 0.65 \pm 0.12$) than in single profile 
BCGs ($\alpha_{single} = 0.59 \pm 0.14$). The mean isophote outer ellipticity (calculated 
at $\mu \sim$ 24 $mag$ $arcsec^{-2}$) is higher in double profile BCGs ($e_{double} = 0.30\pm0.10$) 
than in single profile BCGs ($e_{single} = 0.26\pm0.11$). Similarly, the mean absolute value of 
inner minus outer ellipticity is also higher in double profile BCGs

On the other hand, we did not find differences between these two BCGs categories when we compared 
global cluster properties such as the BCG-projected position relative to the cluster X-ray center 
emission, X-ray luminosity, or BCG orientation with respect to the cluster position angle.

\end{abstract}

\keywords{galaxies: clusters: general --- galaxies: elliptical and lenticular}

\maketitle


\section{Introduction}\label{Intro}

Brightest cluster galaxies (BCGs) are the most massive and most luminous galaxies in the universe 
today. BCGs are so massive ($M_{star}>10^{11} M_{\odot}$) that their formation and evolution is closely tied 
to the large scale structure of the universe (Conroy et al. 2007). Semi-analytical models show how 
BCGs are formed through complex hierarchical merging of many small galaxies and originate within 
the densest dark matter halos of primordial density fluctuations. BCGs reside only in overdense 
regions of the Universe such as galaxy clusters and groups where merging occurs at a high rate 
over cosmic time (de Lucia \& Blaizot 2007). It is precisely the accretion of numerous stellar 
systems that give the BCGs their apparently homogenous properties. For instance, their total luminosities 
are relatively constant and can be used as standard candles (Lauer \& Postman 1992).

For several decades the luminosity profiles of elliptical galaxies were modeled with the empirical 
$R^{1/4}$ de Vaucouleurs law (de Vaucouleurs 1948). However Lugger (1984) and Schombert (1986) showed 
that most elliptical galaxies have a flux excess at large radii with respect to the $R^{1/4}$ model. 
Schombert (1987) modelled the BCG light profiles with a power law rather than the de Vaucouleurs 
law underscoring the different nature of BCGs and standard elliptical galaxies.

A model used virtually by every author in recent years to fit the light profiles of a wide range 
of stellar systems is the generalization of the de Vaucouleurs law introduced by S\'ersic (1963, 1968). 
The S\'ersic model in the form $R^{1/n}$  is a mixture of bulge and disk components using only three free 
parameters ($\mu_{e}$, $r_{e}$, and $n_s$) instead of four ($\mu_{e}$, $r_{e}$, $\mu_0$, and $r_0$) (see 
Section 3 and the comprehensive review of Graham and Driver 2005). 

More recently, several papers have suggested that a simple S\'ersic $R^{1/n}$ law does not properly model 
the luminosity profile of some elliptical (usually cD) galaxies. Gonzalez et al. (2003, 2005)
found that the best fit to the light profiles of 30 BCGs was a double $R^{1/4}$ de Vaucouleurs 
profile.  Seigar et al. (2007) studied the light profiles of five cD galaxies and showed that 
a S\'ersic plus an exponential function are necessary to accurately reproduce an inner and an outer 
component present in their surface brightness profiles. Donzelli et al. (2007) estimated that roughly half of 
82 elliptical galaxies belonging to the 3CR radio catalog also require a S\'ersic + exponential model 
to properly fit their light profiles. Using numerical simulations Hopkins et al. (2009) propose that
dissipational mergers are at the origin of the double components  light profiles in the core of elliptical 
galaxies. According to their models a violent relaxation of stars whose parent galaxies
participate in the merger is responsible for the creation of an outer component while a central 
starburst gives rise to the inner component.

We use a homogeneous and uniquely large sample of ground-based imaging of Abell clusters
to carefully examine the luminosity profiles of 430 BCGs and determine the best fitting function
and structural parameters. This is key to properly constrain dynamical models and the merging 
history of these galaxies.

The paper is organized as follows. In Section 2, we present the observations and reductions, 
while in Section 3 we describe data processing. Light profile fitting procedure and structural parameters
are discussed in Section 4, while in Section 5 we discuss the results and in Section 6 we summarize the conclusions.


\section{Observations}\label{observations}

BCG images used in this work were provided by M. Postman (STScI) who kindly gave us access to the raw data.
They were obtained under photometric conditions using the KPNO 2.1 m and 4 m telescopes,
and the CTIO 1.5 m telescope between 1989 November and 1995 April over a total of 13 observing runs
(see Table 1). Five of these runs are described in more detail in Postman \& Lauer (1995).
Briefly, all the images were acquired in the Kron-Cousins $R_c$ band and have typically exposure times
of 200-600 s. During these runs seeing conditions were very good to fair, namely, FWHM = 1"-2", and 
nights were photometric.\\
In order to flat field the images a series of dome flats were obtained each night. This allowed for 
flatfielding with a typical accuracy better than 0.5\% of the sky level. 
Photometric calibration was obtained by observing 10-15 Landolt (1983) standard
stars per night. This also enabled us to calculate extinction coefficients and to check the zero point
on each night of the observing runs. The overall photometric accuracy was better than
0.02 mag, much smaller than the typical errors of the BCG photometric parameters, which
are more sensitive to background subtraction, and fitting models.

One of the key points of this homogeneous sample is that approximately 50 BCGs were observed in 
common between the different runs. Many of these BCGs were actually observed in five or more runs. 
This not only allowed us to verify and compare reductions for all observing runs, but also to improve 
the accuracy of the luminosity profiles as discussed in the next section. The rms scatter for the
integrated magnitudes of the galaxies is 0.03 mag, while for the luminosity profiles at
$\mu_{R} =$ 22.5 $mag$ $arcsec^{-2}$ is 0.11 $mag$ $arcsec^{-2}$.

\section{Data Processing and Sky Level}\label{DP}

All images were processed following the standard recipes: bias and flat field corrections using 
IRAF routines. After this process we carefully inspect the images in order to determine the 
best method to subtract the sky background. Gauging the sky is a crucial step since sky subtraction 
has a significant influence on the faint end of the luminosity profiles and therefore 
on the structural parameters we derive. In most cases a two-dimensional first-degree polynomial was 
sufficient to give an accurate fit to the sky, and we used the residuals distribution to estimate 
the uncertainty of the sky level $\sigma_{sky}$. The importance of sky cleaning is summarized 
in detail in Coenda et al. (2005).

Similar tests were made to measure the effect of seeing on the luminosity profiles.  
The effects of seeing dictate that the minimum radius for a suitable fit to 
the luminosity profile is $r=1.5\times$FWHM, this is particularly true for large galaxies, i.e.,
apparent radius greater than  $\sim12\times$FWHM (Coenda et al. 2005). We show in detail in the 
next section that BCG galaxies were relatively bright and extended and in most cases 
the apparent radius of the galaxy was greater than $20\arcsec$.

\section{Luminosity Profiles  and Profile Fitting}\label{PF}

We use the  {\sc ellipse} routine (Jedrzejeswki, 1987) within Space Telescope Science 
Analysis System (STSDAS) to extract the luminosity profiles of the BCGs. We apply this routine to the 
processed, sky subtracted images. In many cases galaxy overlapping is an issue due to 
the crowded fields around BCGs. To solve this, we apply the technique described in Coenda et al. 
(2005) which consists in masking the overlapping regions before profile extraction. Then we 
obtain the luminosity profile and structural parameters (center coordinates, ellipticity, and 
position angle of the isophotes), and construct a model BCG galaxy that is subtracted from the
original image. The residual image is then used to extract the luminosity profile of the 
overlapping galaxies. The process is repeated several times until the profile of the 
target galaxy converges.

Isophote fitting was performed down to a count level of 2$\sigma_{sky}$; i.e., the
fitting procedure was stopped when the isophote level was around twice the background
dispersion (pixel-to-pixel variance), which in our case corresponds to surface
magnitudes between $\mu_{R} \sim$ 23.5 - 24.5 $mag$ $arcsec^{-2}$,
depending on the observational run when the cluster was observed.

For each cluster, we usually obtain the luminosity profiles for the three brightest galaxies
in the field. This preliminary selection is made by eye. In those cases where the 
selection is not obvious, we also obtain additional luminosity profiles, i.e., for the five 
brightest galaxies. The final BCG selection is made through the galaxy metric luminosity, that 
is the luminosity enclosed in a radius of 14.5 kpc, we also included galaxy redshift to ensure 
cluster membership. Redshift data were obtained from the NASA/IPAC Extragalactic Database (NED).

As discussed in the introduction, there are a wide variety of functions to perform 
luminosity profile fitting. We adopted the S\'ersic profile $R^\beta$, where the 
concentration parameter $\beta = 1/n$ is the inverse of the S\'ersic index (S\'ersic 1968):

\begin{equation} \label{Ser}
I(r)=I_{e} exp\Big\{-b_n\Big[\Big(\frac{r}r_{e}\Big)^{\beta} - 1\Big]\Big\}.
\end{equation}

In this equation $I_{e}$ is the intensity at $r = r_{e}$ the radius that encloses half of 
the total luminosity (also known as the effective radius or half-light radius). In this context 
$b_n$ can be calculated using $b_n \sim 2\/n-0.33$ (Caon et al. 1993).  

We use the {\sc nfit1d} routine within STSDAS to find the coefficients that best 
fit the light profiles of each galaxy. This task uses a $\chi^2$ minimization scheme to
fit the best non linear functions to the light profiles tables we obtained with 
{\sc ellipse} (Schombert \& Bothun 1987). The fitting procedure is carried out only 
in the portion of the galaxy surface brightness profile where the signal-to-noise ratio 
was greater than 3 ($S/N > 3$). We did this in order to avoid the regions at the faint 
end of the luminosity profiles, in which errors are large. Moreover, we have also excluded 
the inner 3"-4" of the luminosity profiles to avoid the blurring effects of seeing
(see the test described in Coenda et al. 2005). 

Errors on the structural parameters were calculated following the method described by Coenda 
et al. (2005). Briefly, this technique consists of creating test images with model galaxies
that have known luminosity profile parameters. Then we artificially add and subtract to those images 
a constant value corresponding to $\sigma_{sky}$. Finally we extract and refit the new 
luminosity profile as explained above. These newly obtained parameters give us the respective 
upper and lower limits.

To a first approximation a single S\'ersic model provides a good fit to the light profile of 
our sample of BCGs, as shown by Graham et al. (1996). However, for almost half of the sample we 
noticed that a single S\'ersic function fails to properly reproduce the BCGs surface brightness 
in the range $\sim22.0$ $mag$ $arcsec^{-2}$.  Note that this case is very similar to that 
presented in La Barbera et al. (2008, their Figure 18). It is evident that the 
S\'ersic function do not properly fit the luminosity profiles of these elliptical 
galaxies in the inner 4$\arcsec$, where the residuals are almost $\sim0.3$ $mag$ $arcsec^{-2}$.

It is not necessary to have very deep luminosity profiles, as in the case of Seigar et al. (2007), 
to realize that for certain galaxies, even in regions of bright surface luminosity,  a single 
S\'ersic model cannot account for the concave shape of the luminosity profile.

For these galaxies their light profiles were best fitted adding to the S\'ersic model of Equation
(1) an outer exponential function (Freeman, 1970):

\begin{equation} \label{exp}
I(r)= I_{0} exp\Big(-\frac{r}r_{0}\Big).
\end{equation}

In this equation $I_{0}$ is the central intensity  and $r_{0}$ is the
length scale. The inclusion of this equation in the fitting function does not necessarily 
means that the galaxy has a disk component in the usual sense of rotation-supported stellar system. 

We have chosen the exponential function since it is the simplest function to account
for the "extra-light" observed in the above mentioned galaxies. It is worth to mention that we also tried with a
second S\'ersic function since it has three degrees of freedom instead two. However, in terms of the $\chi^2$
coefficient it is not better than the S\'ersic plus exponential fit. 
Figure 1 shows a good example:
the case of BCG A0690 where clearly a single S\'ersic function cannot account for the concavity
of the luminosity profile. Even when the fitting becomes adequate at the faint end of the profile,
the model cannot reproduce the surface magnitude in the interval between 7 and 15 kpc, 
which corresponds to a surface magnitude in the range of 21-22 $mag$ $arcsec^{-2}$. 
Error bars in this region have approximately the size of the squares. 
On the other hand, Figure 2 shows a much better fit due to the inclusion
of the exponential component (dashed line) in the fitting model. We show that 
the 21-22 $mag$ $arcsec^{-2}$ region is now in perfect agreement with the model and the fitting 
functions can also properly describe  the faint end of the luminosity profile. In fact, 
the reduced $\chi^2$ coefficient we obtain with a S\'ersic + Exponential model 
is about one order of magnitude smaller than those obtained with the single S\'ersic fit.
As a general rule, for most of those cases that are initially fitted with the 
single S\'ersic function and in which $n > 8$ and $r_{e} > 300$ kpc, it is necessary 
to include the exponential component to obtain a proper fit.

Intensity  parameters  are  then  converted into  surface  brightness, expressed in  
$mag$ $arcsec^{-2}$ by the  equation $m=-2.5log(I)$, while units of $r_e$,  and $r_0$ 
are converted to  kpc. Errors for $r_e$ and $r_0$  are smaller than 15\% while for  
$\mu_e$, and $\mu_0$ errors are below 0.20 $mag$ $arcsec^{-2}$. Total luminosities of 
both S\'ersic and the exponential components are finally computed using the derived
photometric  parameters  and  integrating  separately  both  components  of
Equations (1) and (2) as follows:

\begin{equation}
\label{luminosidad}
L=\int_{0}^{\infty}I(r)2\pi rdr,
\end{equation}
which yields
\begin{equation}
L_{Sersic}=I_{e}r_{e}^{2}\pi\frac{2exp(b)}{\beta k^{2/\beta}} \Gamma(2/\beta) 
\end{equation}
for the S\'ersic component and
\begin{equation}
L_{exp}=2\pi I_{0}r_{0}^{2}
\end{equation}
for the exponential component. $\Gamma(2/\beta)$ is the gamma
function. Total apparent magnitudes
are then  converted into absolute magnitudes.   Throughout this paper
we  assume a  Hubble  constant  $H_0$ =  70  $km$ $s^{-1}$  $Mpc^{-1}$
together with $\Omega_M$ = 0.27 and $\Omega_{\Lambda}$ = 0.73.\\


\section{Results and Discussion}\label{RD}

Table 2 lists  the   photometric  parameters  obtained  through  the fitting procedure described 
in Section 3 for all BCGs of the sample. Columns  1-6 list  the  Abell BCG name, S\'ersic 
parameters $\mu_{e}$,  $r_{e}$, $n$, exponential parameters $\mu_0$  and  $r_0$. Columns 7-10 give
absolute magnitude of the S\'ersic component, absolute magnitude of the exponential
component, total absolute magnitude of the BCG and S\'ersic to exponential component light ratio.
Column 11 contains the logarithmic slope of the metric luminosity or $\alpha$ parameter which is
defined as $\alpha = d(Log(L_m))/d(log(r_m))$, where $L_m$ is the total BCG luminosity
within a circular aperture of radius $r_m$ centered on the BCG nucleus. Following Postman \& Lauer (1995)
we have calculated this parameter at r = 14.5 kpc. Columns 12-15 list the inner ellipticity 
(measured at 10$\arcsec$), outer ellipticity (measured at $\sim23-24$ $mag$ $arcsec^{-2}$), 
and inner and outer position angles of the isophotes where the ellipticities are measured. 
Position angles are from north to east and the typical observed errors are $\sim5^o$, 
while typical errors for ellipticities are 0.06. Finally, column 16 lists the metric absolute 
magnitude also calculated at r = 14.5 kpc.

The data above show that 225 out of 430 BCGs, or 52\% of the sample, have a single S\'ersic 
luminosity profile, while the remaining 205 BCGs (48\%) need a double component of S\'ersic + exponential
to properly fit their luminosity profiles. We note  that we find 27 galaxies 
($\sim6\%$ of the whole sample) that have $n < 1.5$ for the inner S\'ersic component, which is nearer to an
exponential rather than a de Vaucoleurs profile ($n \sim 4$). Moreover, all but three, 
have double component luminosity profiles. This is particularly interesting since it has been suggested 
that BCGs usually have high S\'ersic indices (Graham et al. 1996). However, Seigar et al. (2007) observe
an inner exponential behavior in 3 out of 5 galaxies of their sample, suggesting that this may be more
common in cD galaxies than previously thought. A visual inspection of these galaxies reveals that
they have high ellipticities. We calculated an average outer ellipticity $e = 0.32\pm0.09$, which is
slightly higher than the average found for double profile BCGs (see the next section).

In general terms we have also noticed that the inner components of the double profile BCGs have  
effective radii $r_e \sim 1-10$ kpc and S\'ersic indices $n \sim1-6$, being the averages 
$r_e = 5 \pm4$ kpc and $n = 3.7 \pm1.5$ respectively. This values are quite similar to those 
reported by Gonzalez et al. (2003) in their preliminary paper for a sample of 31 BCGs. However, these 
authors use a different approach to fit the luminosity profiles. They use two S\'ersic functions 
instead our S\'ersic + exponential approach.


\subsection{Single Profile BCGs versus Double Profile BCGs}

A question that arises after our analysis is to establish if single profile BCGs actually differ 
in morphology from double profile BCGs. The models of the light profiles we apply do not imply
conjectures on galaxy morphologies. Are BCGs with single and double profile of a different 
nature? If so, is this difference environmental or intrinsic?

In order to answer these questions we carried out a series of tests in which
we explored BCGs properties together with the global cluster properties.
One of them is the Kormendy relation (KR; Kormendy 1977) which is presented in Figure 3.
This is an empirical scaling relation between surface brightness $\mu_e$ and effective radius
$r_e$ and it represents the projection of the Fundamental Plane (Djorgovski \& Davies 1987).
For the case of single profile BCGs both $\mu_e$ and $r_e$ are directly obtained from the fitting
profile. However, in the case of double profile BCGs, we calculate these 
parameters from the double profile, i.e., using the sum of the S\'ersic and exponential profiles.
A linear  regression applied to the whole sample gives\\

\begin{center}
$\mu_e$ = 18.72($\pm0.06$) + 3.13($\pm0.05$) log($r_e$/kpc)\\
\end{center}

The slope of the KR obtained for all BCGs is $a_{BCG} = 3.13 \pm 0.05$ which is similar to 
the value obtained by Oegerle \& Hoesel (1991) for a sample of 43 BCGs (i.e., $a_{BCG} = 3.12 \pm 0.14$). 

However, Bildfell et al. (2008) obtain for a sample of BCGs selected from 48  X-ray luminous clusters 
($a_{BCG} = 3.44 \pm 0.13$), which is considerably steeper than our value and than the value of
"normal" ellipticals $a_{ellip} = 3.02 \pm 0.14$ (Oegerle \& Hoesel, 1991). 
Nevertheless, when we apply the same analysis to single profile and double profile BCGs
separately we obtain for single profile BCGs
\begin{center}
$\mu_e$ = 18.65($\pm0.07$) + 3.29($\pm0.06$) log($r_e$/kpc)\\
\end{center}
and for double profile BCGs
\begin{center}
$\mu_e$ = 19.03($\pm0.10$) + 2.79($\pm0.08$) log($r_e$/kpc).\\
\end{center}

Different slopes suggest that the formation timescale of the 
single profile BCGs could differ from their double profile counterpart (von der Linden et al. 2007).

It is interesting to note that these values are calculated integrating the galaxy luminosity 
profiles up to infinity. If we consider a finite radius instead, total luminosities will 
change and therefore both $r_e$ and $\mu_e$ will also change. We then have also calculated the KR for 
both subsamples considering different galaxy radii ($r = 100, 200, 300$ kpc). We observed that the slope 
of the KR flattens for smaller radii and tends to a similar value for both subsamples ($a\sim2.6$ at $r = 100$ kpc). 
This effect could easily explain the differences found in the size-luminosity relation between von der Linden et al. (2007)
and Lauer et al. (2007) and Bernardi et al. (2007). von der Linden (2007) defines the $r_e$ by integrating luminosity profiles 
up to the $\mu = 23$ $mag$ $arcsec^{-2}$ isophote, while Lauer et al. (2007) and  Bernardi et al. (2007) integrate
the luminosity profiles up to infinity. 

We also find that single profile BCGs show a median total absolute magnitude $M_{T, single} = -23.8 \pm 0.7$, 
while double profile BCGs have $M_{T, double} = -24.0 \pm 0.5$. The  Kolmogorov-Smirnov (K-S) test applied 
to these data indicates that the $M_T$ distributions for single profile and double profile BCGs are 
statistically different at 99.4\% confidence level. Figure 4 shows the total absolute magnitude 
distributions for single and double profile BCGs.

To verify our findings and to rule out a possible dependence on our fitting models, 
we calculated the total luminosity within different diaphragms with radius ranging from 5 to 70 kpc. 
We find that integrated luminosities for both subsamples are indistinguishable 
up to $r = 15$ kpc. This can be seen in Figure 5 where we plot the absolute integrated magnitude 
versus the radius of the circular diaphragm expressed in kpc. The vertical line indicates 
$r = 14.5$ kpc where metric luminosity and alpha parameter are calculated. Average integrated 
luminosities beyond 20 kpc are $\sim 0.2$ mag brighter for double profile BCGs.
We have applied the K-S test to both subsamples and results indicate that they do not statistically
differ for $r =$ 5, 10, and 14.5 kpc, while for larger radii the integrated luminosity distribution
are truly different at the 99 \% confidence level.
We highlight that $r = 20$ kpc is close to value for which the S\'ersic component equals 
the exponential component (see the case for A0690 in Figure 2).
From the data tabulated in Table 2, it is straightforward to compute the radius where 
$I_{Sersic}$ = $I_{exp}$ for each of the double profile BCGs. 
Averaging these values we find $<r> = 13 \pm 5$ kpc at $<\mu> = 22.5 \pm 0.7$ $mag$ $arcsec^{-2}$.
In other words, this result corroborates that the extra-light observed in double profile BCGs 
originates in the intermediate regions of the galaxies and not in the inner regions.
The same conclusion can also be derived from Figure 6 where we plotted the total sum of the 
luminosity profiles corresponding to the single and double profile BCGs. Prior to the sum, 
individual profiles are normalized to their effective radii. Note that the extra-light
of the double profile BCGs becomes apparent in the region
1 $ < r/r_e < $5, which roughly correspond to $\sim15-75$ kpc.


\subsection{Ellipticities and Isophote Twisting}

We have also explored other photometric parameters which are not directly 
related to the surface brightness profile fitting functions. Figure 7 shows the ellipticity 
distributions for single and double profile BCGs, and Figure 8 shows the absolute value of the inner minus outer 
ellipticity distributions for the same subsamples. Outer ellipticities are measured at $\mu \sim$ 23-24 
$mag$ $arcsec^{-2}$ which is approximately half a magnitude brighter than our limit 
surface magnitudes. Similarly, the inner ellipticity is
measured at $r \sim 4-5\arcsec$, which corresponds to 3-4 times the average seeing.
In Figure 7 we show that double profile BCGs have higher ellipticities that single profile BCGs. 
We obtain an average ellipticity $<e_{double}> = 0.30 \pm 0.10$ for double profile BCGs and 
$<e_{single}> = 0.26 \pm 0.11$ for single profile BCGs. A K-S test
applied to these data indicates that the ellipticity distributions for single and double profile BCGs are 
statistically different at a 98.3\% confidence level. Similar results are obtained for the inner
minus outer ellipticity of these subsamples. We obtain an average $< \Delta e_{double} > = 0.15 \pm 0.10$ for
the double profile BCGs, while for single profile BCGs we have $< \Delta e_{single} > = 0.10 \pm 0.09$. 
Again, the K-S test indicates different distributions at 99.9\% confidence level. 
The logarithmic slope of the metric luminosity (calculated at $r = 14.5$ kpc) $\alpha$ is 
also higher in double profile BCGs ($< \alpha_{double} > = 0.65\pm0.12$) than in single profile BCGs 
($< \alpha_{single} > = 0.59\pm0.14$). Figure 9 presents the distributions of $\alpha$ for single 
and double profile BCGs respectively.  A K-S test applied to these data establishes that the 
distributions are statistically different at a 99.9\% confidence level. 

The presence of isophote twisting was also explored. We calculated the outer minus inner 
position angle of the isophotes for those  galaxies with ellipticities greater than 0.15 
since position angle errors associated with rounder isophotes are large. We found similar values 
for both single and double profile BCGs ($< \Delta pa > = 8^o \pm9^o$).


\subsection{S\'ersic + Exponential or Exponential + S\'ersic?}

By combining a large set of hydrodynamical simulations spanning a broad range of luminosity profiles 
of various masses, Hopkins et al. (2009) show an alternative way to separate luminosity 
profiles into an inner starburst component and outer pre-starburst component for "cusp" ellipticals
which are formed via gas-rich mergers. These authors show that dissipational mergers give rise to 
two-component luminosity profiles which can be accounted 
by an exponential function (inner component) plus a S\'ersic model (outer component). The exponential
function accounts for the extra-light that was formed in a compact central starburst and makes the 
inner light profile of the galaxy deviate from a single S\'ersic in the galaxy core. The outer 
component was formed by violent relaxation of the stars already present in the precursor galaxies.

We fitted to our double component BCGs an exponential (inner) + 
S\'ersic (outer) model in the order proposed by Hopkins et al. (2009). As an example, we show the 
results of this fitting for the BCG A0690 in Figure 10. The agreement between model and measurements is 
excellent and the Hopkins model properly accounts for the luminosity profile. 
We also compared the rms and $\chi^2$ values with those obtained 
with our original S\'ersic + exponential model and we find that both models are equally good. In 
other words, our approach and the Hopkins model are, from a mathematical point of view, equivalent. 

The results of fitting exponential +  S\'ersic models to our double 
profile galaxies are summarized in Table 3, which lists the same 10 parameters as in Table 2. 
In this case S\'ersic parameters $\mu_{e}$, $r_{e}$ and $n$ correspond to the outer component, 
while the exponential $\mu_0$ and $r_0$ parameters correspond to the inner component. Obtained 
$\chi^2$ values indicate that this fit proved to be equally good as the inner S\'ersic + outer 
exponential form. However, for some galaxies we inexorably obtained unrealistic ($\geq 300$ kpc) 
effective radius using the Hopkins model, this is not the case for the S\'ersic + exponential 
approach. Given that our results suggest that the extra-light comes from the intermediate 
regions (see also the next section), we believe that the S\'ersic + exponential profile fitting is 
the appropriate selection for the BCGs analized in the present work.


\subsection{Extra-light and D-cD Envelopes}

Around $\sim20\%$ of giant elliptical galaxies have extensive, low-luminosity envelopes.
These galaxies are known as D type, and those with the largest envelopes are denominated 
cD galaxies (Mackie, 1992). The envelopes, which are seen as deviations from the de 
Vaucouleurs profile, are quite faint, occur below 24 $mag$ $arcsec^{-2}$ in the $V$ band, 
and they extend well beyond 100 kpc in projected size. Thus, only few giant ellipticals 
have confirmed envelopes. 

We explore if the extra-light found at intermediate radii is related to an eventual 
cD envelope. From the works of Kemp et al. (2007), Seigar et al. (2007), Mackie (1992), 
and Schombert (1986, 1987, 1988), we have found 24 BCGs, cataloged as cD galaxies with 
confirmed envelopes, in common with our sample. These are: A85, A150, A151, A193, A262, A358, 
A539, A779, A1177, A1238, A1767, A1795, A1809, A1904, A1913, A2028, A2147, A2162, A2199, A2366, 
A2572, A2589, A2634,and A2670. Table 2 shows that 19 (79\%) of these 24 galaxies are double 
profile BCGs, while 5 (21\%) are single profile BCGs. Note that three galaxies (A151, A1767, and A2589) 
from those five single profile BCGs, have $r_e > 100$ kpc, and $n > 6$. These parameters are very 
close (see Section 4) to the limit ($r_e > 300$ kpc, $n > 8$)  where it is necessary to include 
an additional exponential profile to obtain a reasonable fit.
On the other hand, one must also note that a visual classification as cD, does not necessarily 
imply the presence of an envelope. Schombert (1986), and Seigar et al.\ (2007) cataloged A496, 
A505, A1691, A2029, A2052, A2107, A2197, and A2666, as cDs without envelopes. For all those galaxies, 
except A2197, our surface profile modeling is consistent with just one component.

These results strongly suggest that the extra-light found in double component BCGs at 
intermediate radii is related to the faint envelope. Moreover, they indicate that this component 
is not only confined to the outskirt of the parent galaxy. Galaxy halos generally considered
an outer component in galaxy structure appear to originate in the inner regions of BCGs.


\subsection{BCGs Morphologies and IR Emission}

Although a visual inspection of our BCGs suggests that they all are early-type galaxies, we
probe the possibility that the differences we observe in the profiles are due to actual
differences in the morphology. The Galaxy Zoo project (http://zoo1.galaxyzoo.org/) provides
the morphological types for a large sample of galaxies observed by the Sloan Digital Sky Survey 
(SDSS; Lintott et al. 2010). In this catalog 66 of the BCGs studied here have a morphological classification:
31 (47\%) are single profile BCGs, while 35 (53\%) are double profile BCGs.
The Lintott et al. (2010) catalog gives the probability that a particular
galaxy is an early-type galaxy or late-type (spiral) galaxy. We find that both single and double
profile BCGs have the same probability to be classified as elliptical galaxies in the Galaxy
Zoo catalog. However, from the 31 single profile BCGs with morphological classification
only 5 (16\%) belong to the 'unknown' category, while for the 35 double profile BCGs in the catalog
this number rises up to 9 (26\%). It is also interesting to note that a visual inspection of these
nine double profile BCGs with unknown morphology are mostly interacting galaxies with
two or three near companions, while the five single profile BCGs with unkown classification
appear to us as normal ellipticals. 


We have scrutinized the infrared emission of a subsample of our BCGs using data from Quillen et al. (2008). 
They report on an imaging survey with the $Spitzer Space Telescope$ of 62 BCGs with optical line 
emission. We have 12 BCGs in common with Quillen et al. (2008), 6 having single component luminosity 
profiles and 6 having double component luminosity profiles. Analysis of the 24-8 $\mu$m flux 
ratios shows that only one (17\%) of single profile BCG (A2052) have infrared excess, 
while for double profile BCGs this number increases to 4 (67\%). Infrared excess is a 
star formation signature. In fact, O'Dea et al. (2010) in their study of seven BCGs using
$Hubble Space Telescope (HST)$ ultra-violet 
and $Spitzer$ infrared data found that all these galaxies have extended UV continuum and
$L-\alpha$ emission as well as an infrared excess. Based on their findings O'Dea et al. (2010)
confirm that the BCGs they study are actively forming stars.
Moreover, they suggest that the IR excess is indeed associated with star formation and they also 
confirm that the FUV continuum emission extends over a region $\sim$ 7-28 kpc. Although these 
results are only for a few BCGs in our sample, they suggest that the "extra-light" observed in 
the double profile BCGs indicates active star formation in the intermediate regions 
of these galaxies. 


\subsection{Global Cluster Properties}

In this section we compare BCGs properties to those of the host cluster such as cluster 
X-ray luminosity, the projected distance of the BCG with respect to the center of the X-ray emission
and the BCG position angle with respect to that of the cluster.
We have used data taken from Ledlow et al. (2003) to determine the offset in kpc 
between the X-ray peak and the optical position of the BCG. X-ray cluster luminosities
were taken from Sadat et al. (2004) data, while the position angle of the clusters
are obtained from Plionis (1994), Rhee \& Katgert (1987), and Binggeli (1982). In this 
case we have only selected those clusters with ellipticities $>$ 0.15, since for 
smaller ellipticities position angles have large errors.\\

Single and double profile BCGs have similar orientations relative to the whole cluster and
X-ray luminosity distributions. We did not observe any difference between the single profile 
and double profile BCG samples with respect to the cluster position angle and X-ray cluster luminosity.
However, Bildfell et al. (2008) report that brighter BCGs are located closer to the X-ray peak 
emission. In our case, considering that on average double profile BCGs are brighter than single 
profile BCGs we should observe larger offsets in single profile BCGs. Nevertheless, we find 
no significant differences between single profile and double profile BCGs offsets with respect 
to the X-ray center emission of the cluster. Moreover, we do not find any correlation between 
the total absolute magnitude and X-ray offset. Figure 11 shows that there is no obvious 
trend between the total absolute magnitude of both single profile and double profile BCGs.

\section{Conclusions}\label{Conclusions}

We have established the existence of two subpopulations of BCGs
based on their luminosity profiles. We analyze a uniquely large sample of 430 BCGs
and find that 48\% of these galaxies have a light profile that deviates from 
the standard single S\'ersic model. The luminosity profiles of these galaxies are 
in fact better described by a double component model consisting of an inner  S\'ersic
profile and an outer exponential component. The necessity of an outer exponential component 
conveys the presence of extra-light at intermediate radii corresponding to surface magnitudes $\sim 22 mag arcsec^{-2}$. 
We have found strong evidence from a subsample of 24 BCGs that extra-light is closely related to the presence of a faint
envelope. Similarly, from other subsample of 12 BGCs we also found evidence that links extra-light and star formation.

This work highlights the need to cover a large spatial scale when deriving the structural
parameters of large galaxies. Accurate parameters can only be obtained when the entire galaxy
is considered and this often requires the creation of composite light profiles using 
data from different telescopes as clearly illustrated by Kormendy et al. (2009). 
$HST$ detectors provide a superb spatial resolution that has been fundamental for the study
of the deviation from the S\'ersic law of the light profile in the inner regions of 
galaxies i.e. cusps and evacuated cores (Ferrarese et al. 2006). However, $HST$ detectors
do not  provide the field of view necessary to truthfully derive the structural
parameters of galaxies with a light profile that deviates from a single S\'ersic
law at large radii.

\section{Acknowledgments}

We are thankful to M. Postman (STScI) for giving us access to the data used in this study. 
We also wish to thank the anonymous referee for his useful comments which helped to clarify 
and strengthen this paper. This research has made use of the NASA Astrophysics Data System 
Bibliographic services (ADS) and the NASA/IPAC Extragalactic Database (NED) which is 
operated by the Jet Propulsion Laboratory, California Institute of Technology, under 
contract with the National Aeronautics and Space Administration.\\
This work has been partially supported with grants from Consejo Nacional
de Investigaciones Cient\'\i ficas y T\'ecnicas de la Rep\'ublica Argentina
(CONICET), Secretar\'\i a de Ciencia y Tecnolog\'\i a de la Universidad
de C\'ordoba and  Ministerio de Ciencia y Tecnolog\'\i a de C\'ordoba, Argentina.\\



Bernardi, M., Hyde, J.B., Sheth, R.K., Miller, C.J., Nichol, R.C. 2007, AJ, 133, 1741\\

Bildfell, C.; Hoekstra, H.; Babul, A.; Mahadavi, A. 2008, MNRAS, 389, 1637\\

Binggeli, B. 1982, A\&A, 107, 338\\

Caon, N., Capaccioli, M., \& D'Onofrio, M. 1993, MNRAS, 265, 1013\\

Coenda, V., Donzelli, C.J., Muriel, H., Quintana, H., \& Infante, L. 2005, AJ, 129, 1237\\

Conroy, C., Wechsler, R. H., \& Kravtsov, A. V. 2007, ApJ, 668, 826\\

de Lucia, G. \& Blaizot, J. 2007, MNRAS, 375, 2\\

de Vaucouleurs, G. 1948, Ann. Astrophys., 11, 247\\

Djorgovski, S., \& Davis, M. 1987, ApJ, 313, 59\\ 

Donzelli, C.J., Chiaberge, M., Macchetto, F. D., Madrid, J.P., Capetti, A., \& Marchesini, D. 2007, ApJ, 667, 780\\

Ferrarese, L., et al. 2006, ApJS, 164, 334\\

Freeman, K. C. 1970, ApJ, 160, 811\\

Gonzalez, A. H., Zabludoff, A. I., \& Zaritsky, D. 2003, Ap\&SS, 285, 67\\

Gonzalez, A. H., Zabludoff, A. I., \& Zaritsky, D. 2005, ApJ, 618, 195\\

Graham, A.W., \& Driver, S.P. 2005, PASA, 22, 118\\

Graham, A.W., Lauer, T.R., Colless, M. \& Postman, M. 1996, ApJ, 465, 534\\

Hopkins, P.F., Cox, T. J., Dutta, S. N., Hernquist, Lars., Kormendy, J., Lauer, T. R. 2009, ApJS, 181, 135\\

Jedrzejewski, R. 1987, MNRAS, 226, 747\\

Kent, S.M. 1985, ApJS, 59, 115\\

Kent, S.N., Guzm\'an Jim\'enez, V., Ram\'irez Beraud, P., Hern\'andez Ibarra, F.J., \& P\'erez Grana, J.A.
	2007, in IAU Symp. 235, Galaxy Evolution Across the Hubble Time, ed. F. Combes \& J. Palous (Cambridge: Cambridge Univ. Press),  213\\

Kormendy, J. 1977, ApJ,  218, 333 

Kormendy, J., Fisher, D. B., Cornell, M. E., \& Bender, R. 2009, ApJS, 182, 216\\

La Barbera, F., Busarello, G., Merluzzi, P., De La Rosa, I., Coppola, G., \& Haynes, C.P. 2008, PASP, 120, 681\\

Landolt, A.U. 1983, AJ, 88, 853\\

Lauer, T. R., \& Postman, M. 1992, ApJ, 400, L50\\

Lauer, T.R., et al. 2007, ApJ, 662, 808\\

Ledlow, M.J., Voges, W., Owen, F.N., \& Burns, J.O. 2003, AJ, 126, 2740\\

Lintott, C. et al. 2011, MNRAS, 410, 166\\

Lugger, P.M. 1984, ApJ, 286, 106\\

Mackie, G. 1992, ApJ, 400, 65\\

Oegerle, W.R.; Hoessel, J.G. 1991, ApJ, 375, 150\\

O'Dea, K. et al. 2010, ApJ, 719, 1619\\

Plionis, M. 1994, ApJS, 95, 401\\

Postman, M. \& Lauer, T. R. 1995, ApJ, 440, 28\\

Quillen, A.C. et al. 2008, ApJS, 176, 39\\

Rhee, G., \& Katgert, P. 1987, A\&A, 183, 217\\

Sadat, R., Blanchard, A., Kneib, J.P., Mathez, G., Madore, B., \& Mazzarella, J.M. 2004, A\&A, 424, 1097\\

Seigar, M.S., Graham, A.W., Jerjen, H. 2007, MNRAS, 378, 1575\\

Schombert, J.M. 1986, ApJS, 60, 603\\

Schombert, J.M. 1987, ApJS, 64, 643\\

Schombert, J.M. 1988, ApJ, 328, 475\\

Schombert, J.M. \& Bothun, G.D. 1987, AJ, 93, 60\\

S\'ersic, J.L. 1963, Boletin de la Asociaci\'on Argentina de Astronom\'ia, 6, 41\\

S\'ersic, J.L. 1968, Atlas de Galaxias Australes (C\'ordoba: Obs. Astron\'om.)\\

von der Linden, A., Best, P.N., Kauffmann, G., \& White, S. 2007, MNRAS, 379, 867\\



\begin{figure}
\center
\plotone{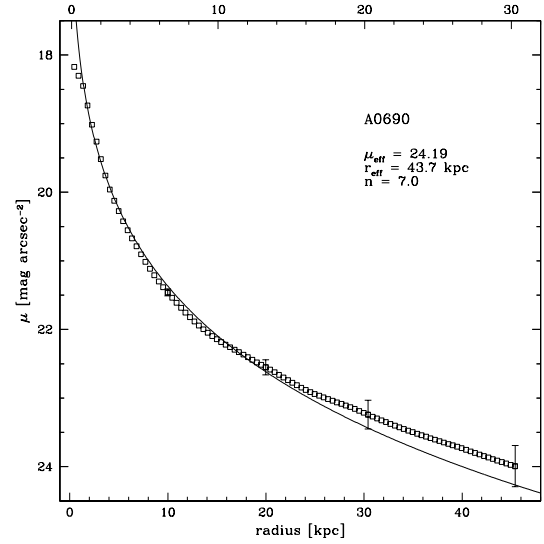}
\caption{
BCG A0690 luminosity profile with the single S\'ersic fitting.
Upper scale is in arcsec. 
}
\label{fig1a}
\end{figure}


\begin{figure}
\center
\plotone{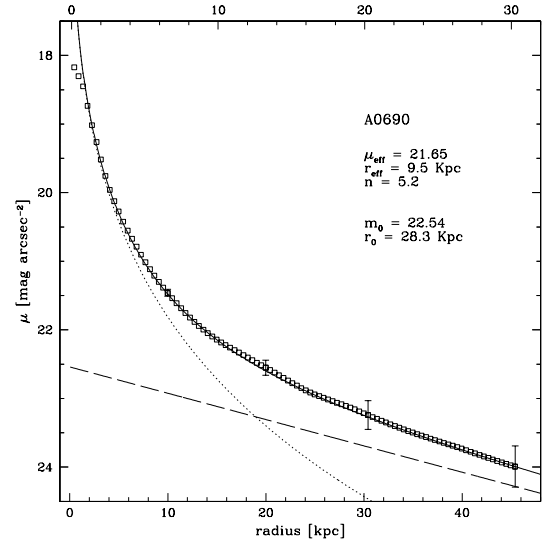}
\caption{
Inner S\'ersic (short dashed line) + outer exponential (long dashed line) fitting model
for A0690 luminosity profile.
}
\label{fig1}
\end{figure}


\begin{figure}
\center
\plotone{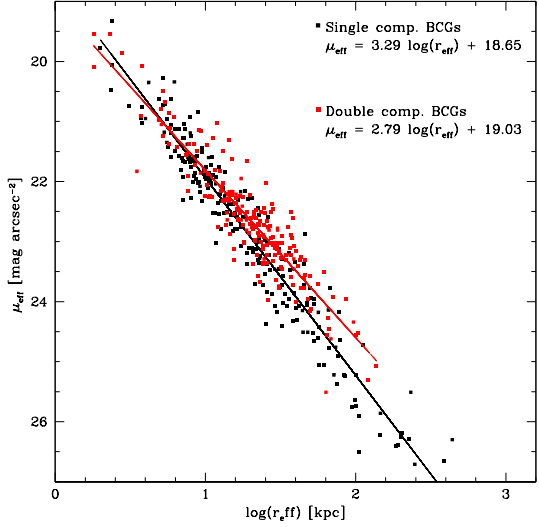}
\caption{
Kormedy relation for sample galaxies. Black dots represent single profile BCGs,
while red dots represent double profile BCGs. Best fits for both subsamples are
also shown. 
}
\label{fig1}
\end{figure}



\begin{figure}
\center
\plotone{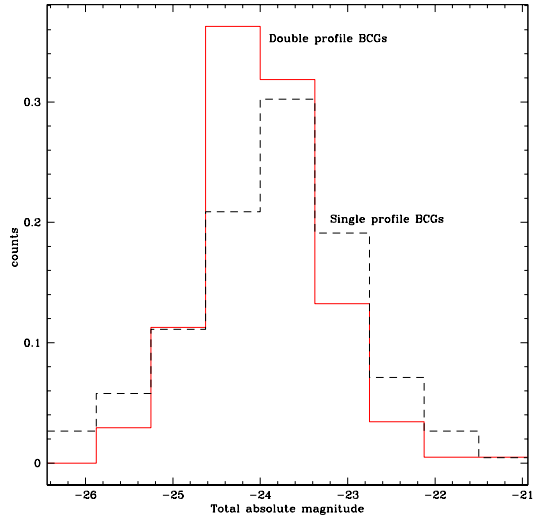}
\caption{
Total magnitude distributions for single (black line) and double profile
(red line) BCGs.
}
\label{fig1}
\end{figure}



\begin{figure}
\center
\plotone{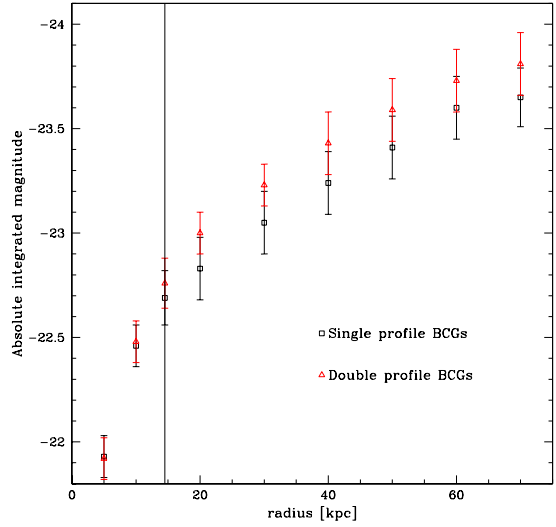}
\caption{
Integrated absolute magnitudes vs. diaphragm radius in kpc. The vertical
line indicates where alpha parameter and metric magnitudes are calculated.
}
\label{fig1}
\end{figure}


\begin{figure}
\center
\plotone{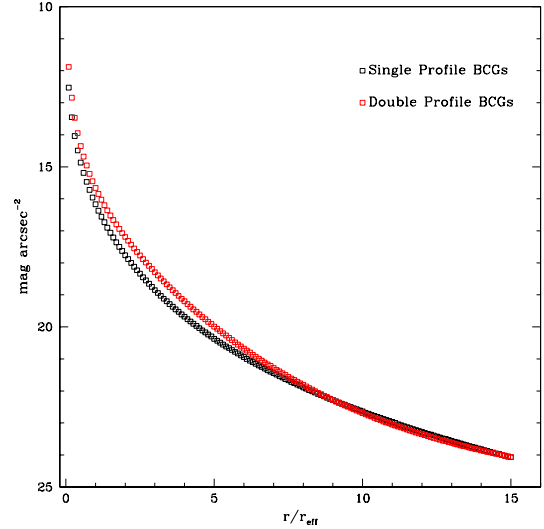}
\caption{
Luminosity profiles obtained using the stacking technique for all single and double profile BCGs.
Prior to the stacking the individual profiles were normalized at the effective radius.
}
\label{fig1}
\end{figure}



\begin{figure}
\center
\plotone{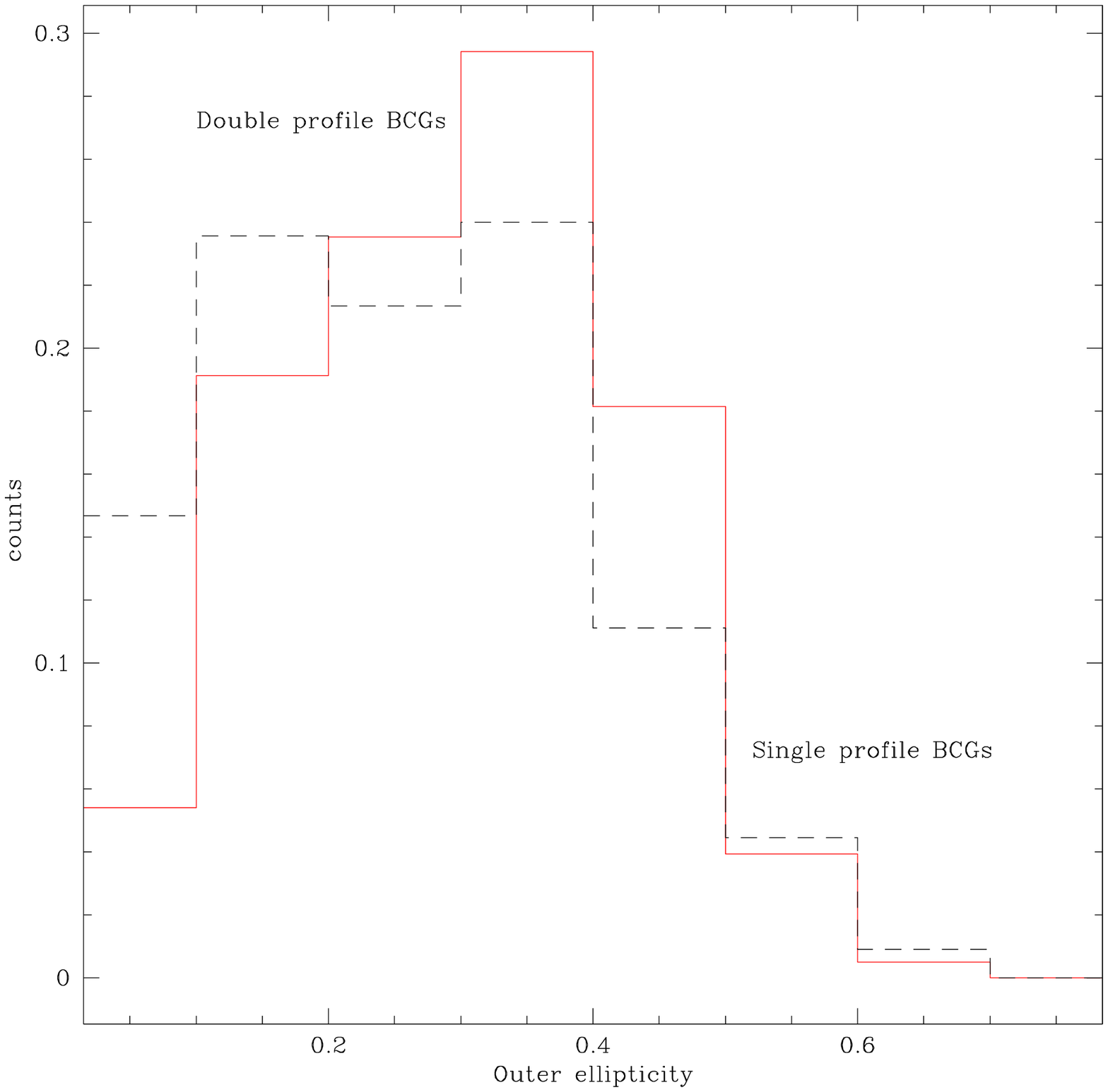}
\caption{
Outer ellipticity distributions for single (black line) and double profile
(red line) BCGs.
}
\label{fig1}
\end{figure}



\begin{figure}
\center
\plotone{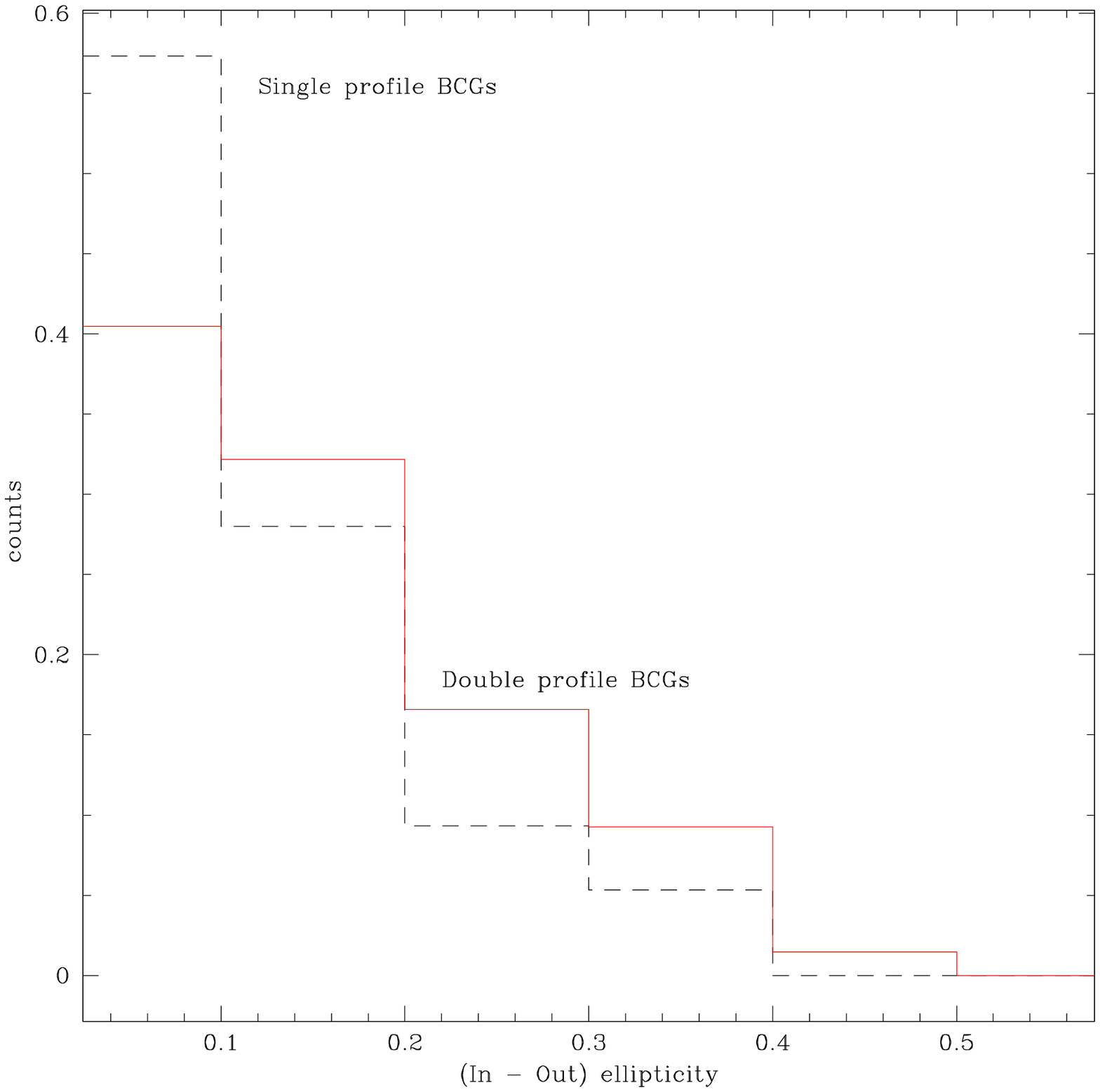}
\caption{
Outer minus inner ellipticity distributions for single (black line) and double profile
(red line) BCGs.
}
\label{fig1}
\end{figure}



\begin{figure}
\center
\plotone{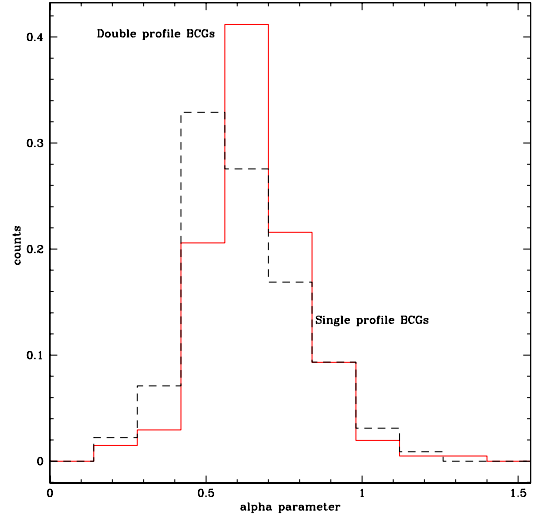}
\caption{
$\alpha$ parameter distributions 
for single (black line) and double profile (red line) BCGs.
}
\label{fig1}
\end{figure}



\begin{figure}
\center
\plotone{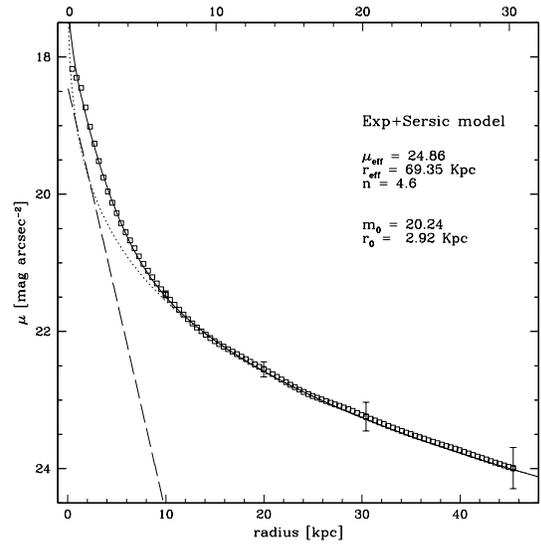}
\caption{
Exp + S\'ersic model fit for A0690 luminosity profile. Outer S\'ersic (short dashed line) and 
inner exponential (long dashed lined) can be observed.
}
\label{fig1}
\end{figure}



\begin{figure}
\center
\plotone{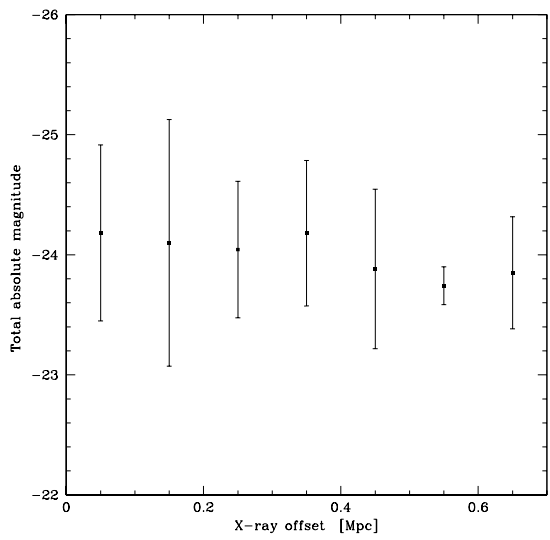}
\caption{
BCGs total absolute magnitude vs. X-ray offset for the whole BCG sample.
}
\label{fig1}
\end{figure}

\clearpage


\begin{deluxetable}{rcccc} 
\tabletypesize{\scriptsize} 
\tablecaption{BCGs Imaging Runs \label{tbl-2}} 
\tablewidth{0pt}
\tablehead{
\colhead{Run} & \colhead{Date} & \colhead{Observatory} & \colhead{Pixel Scale (")} & 
\colhead{FOV (')} 
} 

\startdata
1 & 1989 Oct & CTIO & 0.273 & 3.6 \\ 
2 & 1989 Nov & KPNO 4m & 0.299 & 4.0  \\
3 & 1990 Nov & CTIO & 0.273 & 3.6 \\
4 & 1991 Mar & KPNO 2.1m & 0.304 & 5.2 \\
5 & 1991 Apr & CTIO & 0.434 & 7.4 \\
6 & 1993 Sep & KPNO 2.1m & 0.304 & 5.2 \\
7 & 1993 Nov & CTIO & 0.434 & 7.4 \\
8 & 1994 May & KPNO 2.1m & 0.304 & 5.2 \\
9 & 1994 May & CTIO & 0.434 & 7.4 \\
10 & 1994 Oct & KPNO 2.1m & 0.304 & 5.2 \\
11 & 1994 Dec & CTIO & 0.434 & 7.4 \\
12 & 1995 Apr & CTIO & 0.434 & 7.4 \\
13 & 1995 Apr & KPNO 2.1m & 0.304 & 5.2 \\

\enddata
\end{deluxetable}


\clearpage

\begin{deluxetable}{lccccccccccccccc} 
\tabletypesize{\scriptsize} 
\tablecaption{BCGs Photometrical Parameters \label{tbl-2}} 
\tablewidth{0pt}
\tablehead{
\colhead{Name} & \colhead{$\mu_e$} & \colhead{$r_e$} & \colhead{$n$} & 
\colhead{$\mu_0$} &  \colhead{$r_0$} & \colhead{$M_{Sersic}$} & \colhead{$M_{exp}$} & \colhead{$M_{T}$} & 
\colhead{$S/e$} & \colhead{$\alpha$} & \colhead{$e_{in}$} & \colhead{$e_{out}$}& \colhead{$pa_{in}$}& \colhead{$pa_{out}$}
& \colhead{$M_{metric}$}\\
\colhead{(1)} & \colhead{(2)} & \colhead{(3)} & \colhead{(4)} & \colhead{(5)} &
\colhead{(6)} & \colhead{(7)} & \colhead{(8)} & \colhead{(9)} & \colhead{(10)} & 
\colhead{(11)} & \colhead{(12)}& \colhead{(13)}& \colhead{(14)}& \colhead{(15)}& \colhead{(16)} 
} 

\startdata
A0014 &26.65 &387.58 &5.08 & --- & --- &-26.65 &  0.00 &-26.65 & --- &1.10 &0.35 &0.56 &  56.5 &  52.0 &-22.87\\
A0027 &24.41 & 32.51 &7.58 & --- & --- &-23.67 &  0.00 &-23.67 & --- &0.59 &0.24 &0.28 & -54.0 & -61.2 &-22.18\\
A0071 &20.73 &  5.18 &5.49 & --- & --- &-23.28 &  0.00 &-23.28 & --- &0.49 &0.06 &0.05 &  51.1 &  31.1 &-22.45\\
A0074 &23.24 & 23.04 &6.45 & --- & --- &-24.05 &  0.00 &-24.05 & --- &0.54 &0.28 &0.30 &  42.2 &  39.6 &-22.93\\
A0075 &21.86 & 10.58 &4.74 & --- & --- &-23.55 &  0.00 &-23.55 & --- &0.55 &0.09 &0.23 &  85.0 &  60.4 &-22.74\\
A0076 &24.08 & 34.80 &7.58 & --- & --- &-24.09 &  0.00 &-24.09 & --- &0.57 &0.05 &0.17 & -66.4 &-118.7 &-22.80\\
A0077 &26.39 &190.00 &6.94 & --- & --- &-25.65 &  0.00 &-25.65 & --- &0.91 &0.13 &0.51 & -62.5 & -64.0 &-23.04\\
A0080 &21.59 &  5.92 &2.24 & --- & --- &-22.29 &  0.00 &-22.29 & --- &0.46 &0.11 &0.06 & -64.2 & -71.0 &-21.98\\
\enddata

\tablecomments{Col.  (1),  Abell Cluster;  col (2), effective surface magnitude;  col. (3),
effective radius (kpc); col.  (4), $n$ parameter; col.  (5), central surface magnitude; col. (6), scale length (kpc); col. (7),Sersic absolute magnitude;  
col.  (8), exponential absolute magnitude; col. (9), total absolute magnitude; col. (10), S\'ersic/exponential ratio; col. (11), $\alpha$ parameter; col. (12), inner ellipticity; col. (13), outer ellipticity; col. (14), inner position angle
; col. (15), outer position angle; col. (16), Metric absolute magnitude.}

\end{deluxetable}

\clearpage

\begin{deluxetable}{lccccccccccccccc} 
\tabletypesize{\scriptsize} 
\tablecaption{BCGs Photometrical Parameters (Hopkins model) \label{tbl-2}} 
\tablewidth{0pt}
\tablehead{
\colhead{Name} & \colhead{$\mu_e$} & \colhead{$r_e$} & \colhead{$n$} & 
\colhead{$\mu_0$} &  \colhead{$r_0$} & \colhead{$M_{Sersic}$} & \colhead{$M_{exp}$} & \colhead{$M_{T}$} & 
\colhead{$e/S$} \\
\colhead{(1)} & \colhead{(2)} & \colhead{(3)} & \colhead{(4)} & \colhead{(5)} &
\colhead{(6)} & \colhead{(7)} & \colhead{(8)} & \colhead{(9)} & \colhead{(10)} & 
} 

\startdata
A0085 & 23.72 &   47.0 &  0.6 & 21.42 &  14.3 & -23.91 & -23.84 & -24.63 &  0.935 \\ 
A0133 & 23.23 &   39.0 &  1.2 & 21.04 &   6.9 & -24.29 & -22.66 & -24.51 &  0.222 \\ 
A0150 & 24.34 &   49.2 &  3.5 & 20.84 &   3.3 & -24.22 & -21.25 & -24.29 &  0.065 \\ 
A0152 & 22.77 &   20.5 &  1.4 & 19.63 &   1.7 & -23.54 & -21.07 & -23.65 &  0.102 \\ 
A0193 & 25.18 &  106.4 &  3.5 & 18.99 &   1.9 & -25.04 & -21.86 & -25.10 &  0.053 \\ 
A0194 & 22.05 &   14.2 &  1.7 & 18.39 &   1.1 & -23.31 & -21.09 & -23.44 &  0.130 \\ 
A0208 & 25.42 &   92.5 &  4.6 & 20.24 &   2.5 & -24.87 & -21.43 & -24.91 &  0.042 \\ 
A0257 & 22.64 &   18.5 &  1.9 & 19.72 &   2.1 & -23.68 & -21.58 & -23.82 &  0.144 \\
\enddata

\tablecomments{Col.  (1),  Abell Cluster;  col (2), effective surface magnitude;  col. (3),
effective radius (kpc); col.  (4), $n$ parameter; col.  (5), central surface magnitude; col. (6), scale length (kpc); col. (7),Sersic absolute magnitude;  
col.  (8), exponential absolute magnitude; col. (9), total absolute magnitude; col. (10), exponential/S\'ersic ratio.}

\end{deluxetable}


\end{document}